\documentclass[aps,prb,reprint,amsmath,amssymb,superscriptaddress]{revtex4-2}

\usepackage{graphicx}
\usepackage{dcolumn}
\usepackage{bm}
\usepackage{amsmath}

\usepackage{mathrsfs}
\usepackage{mathtools}
\usepackage{xcolor}
\usepackage{orcidlink}
\usepackage{hyperref}

\begin{document}

\preprint{APS/123-QED}

\title{Confined kinetics and heterogeneous diffusion driven by\\fractional Gaussian noise: A path integral approach}

\author{D.~S. Quevedo\orcidlink{0000-0003-1583-4262}}
\affiliation{Institute for Theoretical Physics, Utrecht University, Princetonplein 5, 3584CC Utrecht, The Netherlands}
\author{F.~S. Abril-Bermudez\orcidlink{0000-0002-2512-4929}}
\affiliation{School of Natural and Computing Sciences, University of Aberdeen, Aberdeen AB24 3UE, United Kingdom}
\affiliation{Department of Physics, Universidad Nacional de Colombia, Bogot\'a, Colombia}
\author{C. Morais Smith\orcidlink{0000-0002-4190-3893}}%
\affiliation{Institute for Theoretical Physics, Utrecht University, Princetonplein 5, 3584CC Utrecht, The Netherlands}

\date{\today}

\begin{abstract}
Many complex systems are described by Langevin-type equations in which the noise exhibits long-range correlations and couples to the system in a state-dependent, multiplicative manner, leading to heterogeneous non-Markovian diffusion. Here, we investigate the problem of diffusion driven by fractional Gaussian noise with a general multiplicative coefficient from a path-integral perspective. Using a stationary-phase approximation, we derive a Gaussian propagator expressed in terms of the Lamperti transform of the process. In the additive limit, our results recover the path-integral representation of fractional Brownian motion based on its Riemann-Liouville formulation and establish its equivalence with the Langevin construction. We further analyze the effect of subordinating the process to a killing rate within the Feynman-Kac framework, and develop a general procedure to derive kinetic equations in terms of effective local Hamiltonians. We show that the interplay between multiplicative diffusion and confinement induces an effective drift term, leading to probability accumulation in regions of low noise amplitude.
\end{abstract}

\maketitle

The fractional Brownian motion (fBm) is a fundamental stochastic model for systems exhibiting long-range dependence and self-affinity~\cite{mandelbrot1985self, gneiting2004stochastic}. It is characterized by the Hurst exponent $0< H <1$~\cite{hurst1951long}, which controls the roughness of the sample paths, the dependence structure of the increments, and the scaling of the mean squared displacement (MSD). In particular, its increments are persistent for $H > 0.5$ and anti-persistent for $H < 0.5$, corresponding to superdiffusive and subdiffusive behavior, with $\mathrm{MSD} \propto  t^{2H}$. In the limiting case $H = 1/2$, the fBm reduces to the classical Brownian motion, with independent increments and normal diffusion, $\mathrm{MSD} \propto t$. 

The first formulation of fBm $B_H(t)$ (known in the literature as Riemann-Liouville definition) was introduced in 1940 by Kolmogorov~\cite{kolmogorov1940wienersche} and L\'evy in 1953~\cite{levy1953random}, who defined it as a generalization of classical Brownian motion $B_{1/2}(t)$ via the Riemann-Liouville fractional integral over the Brownian increments $dB_{1/2}(t)$, weighted by a power-law kernel $t^{H - 1/2}$. This construction was later extended and formalized by Mandelbrot and Van Ness in 1968~\cite{mandelbrot1968fractional}, who introduced an additional term based on Weyl's integral to reduce the excessive influence of the origin. In this framework, the formal time derivative $\xi_H(t) = dB_H/dt$ defines fractional Gaussian noise (fGn), which reduces to white noise in the limiting case $H = 1/2$. This representation naturally leads to a Langevin-type description of fBm, $\dot{x}(t) = \theta_H \,\xi_H(t)$, with $\theta_H$ a generalized diffusion coefficient. 

The three definitions described above yield equivalent MSD and all lead to a Gaussian process. However, this equivalence breaks down when the diffusion coefficient is promoted to a time-dependent random process~\cite{wang2025different}. In that case, the different representations exhibit distinct behaviors: the Langevin formulation exhibits crossover effects in both the MSD and the mean squared increment, the Mandelbrot-Van Ness representation reflects an effective diffusivity governed by its mean value, while the Riemann-Liouville formulation yields nonstationary mean squared increment and autocovariance functions. 

Starting from the Riemann-Liouville definition of fBm, the transition probability of the process can be constructed elegantly from a path-integral perspective~\cite{calvo2008path,sebastian1995path}, relying on the probability measure of white noise, or Wiener measure~\cite{wiener1923differential,wiener1924average}. Using a stationary phase method, the propagator is shown to be Gaussian, with variance $\mathrm{Var}(B_H(t)) \! \propto \! t^{2H}$. This approach, however, relies specifically on the Riemann-Liouville definition of fBm. More recently, Ref.~\cite{meerson2022path} proposes an explicit path-integral representation for fBm and fGn within a broader framework, starting from the Mandelbrot-Van Ness formulation.

The power-law structure of the time correlations of fGn, as well as its stationarity and Gaussianity, make the process particularly suitable for modeling physical systems driven by long-time-correlated noise and for effectively describing nonlinear thermal environments \cite{lutz2001fractional, kupferman2004fractional} and viscoelastic materials \cite{bonfanti2020,joo2023viscoelastic,durang2024generalized, lim2025anomalous}. A significant increase in the use of both fBm and fGn has occurred in the past two decades, together with the widespread adoption of tools from fractional calculus \cite{metzler2004restaurant}. Notably, the inclusion of memory effects and time-delayed responses in the noise correlations gives rise to a broad range of new physical phenomena. Examples of this are: nonergodicity \cite{cherstvy2021inertia} and prethermal phases \cite{quevedo2024emergent} in the presence of inertial effects, divergent persistent diffusion in active particles \cite{gomez2020active,quevedo2025active}, and the break down of the standard structure of stochastic thermodynamics \cite{khadem2022stochastic} and nanoscale engines \cite{squarcini2022fractional}.

Although the use of fGn accounts for non-local effects as memory and time-delayed responses, in many practical situations, the noise couples to the system in a state-dependent fashion, resulting in multiplicative noise that describes heterogeneous diffusion. This form of noise is commonly encountered in financial modeling (e.g., geometric Brownian motion~\cite{giordano2023infinite}, and geometric L\'evy motion~\cite{abril2025path}), models of economic inequality \cite{bouchaud2000wealth, quevedo2020non}, biological dynamics (e.g., population models~\cite{ricciardi1986stochastic}), but also in systems with anomalous heterogeneous diffusion~\cite{cherstvy2013anomalous}. In particular, when heterogeneous diffusion is coupled with fBm, the process exhibits weak ergodicity breaking \cite{wang2020anomalous}, in contrast to additive fBm, which shows ergodic anomalous diffusion.

In this work, we construct a general path-integral representation for the transition probability of heterogeneous diffusion driven by fractional Gaussian noise, $\dot{x}(t) = a(x)\xi_H(t)$, where $a(x)$ is a non-vanishing $\mathcal{C}^1$ function. The action is expressed in terms of the cumulant generating function of the noise and retains a quadratic form, allowing an exact treatment via the stationary-phase approximation. This leads to a simple Gaussian propagator defined through the Lamperti transform of the process~\cite{lamperti1962semi}, which maps the multiplicative dynamics into an additive fBm. Notably, in the additive case $a(x)=\theta_H$, we recover and connect path-integral results based on the Riemann-Liouville formulation of fBm~\cite{calvo2008path,sebastian1995path} with the Langevin representation.

Building on this framework, we develop a general procedure, based on the Feynman-Kac functional~\cite{kac1949distributions,abril2025path}, to derive kinetic equations associated with effective local Hamiltonians after the stationary-phase approximation. This allows us to analyze the differences between local and non-local kinetic equations, as well as the effects of confinement with absorbing boundary conditions. In particular, we show that the interplay between multiplicative diffusion and confinement induces an effective drift term in the kinetic equation, leading to an accumulation of probability in regions of lower noise amplitude. The present formulation can, in principle, be extended to more general non-Markovian dynamics and noise structures~\cite{hanggi1989path,kharchenko2002path}.





%

This paper is structured as follows. In Sec.~\ref{pathintegral}, we introduce the path-integral representation for the multiplicative diffusion driven by fGn. In Sec.~\ref{stationary}, we apply the stationary-phase approximation to obtain exact expressions for the propagator. The analysis of local and non-local kinetic equations, as well as confinement with absorbing boundary conditions, is presented in Sec.~\ref{confinement}. Finally, conclusions and outlook are given in Sec.~\ref{conclusions}.

\section{\label{pathintegral} Path integral representation}

We start by considering the Langevin-type equation for heterogeneous diffusion,
\begin{equation}
    \dot{x}(t) = a(x)\xi_H(t),
    \label{eq:mfBm}
\end{equation}
where $a(x)$ is a multiplicative coefficient dependent of the process $x(t)$ itself and $\xi_H(t)$ is fGn with zero mean $\langle \xi_{H}(t) \rangle=0$, Hurst exponent $0 \! < \! H \! < \! 1$, and continuous-time correlations, for $t \! \neq \! s$,
\begin{equation}
    C_H(|t-s|) = \left\langle \xi_{H}(t)\xi_{H}(s)\right\rangle = H(2H -1) |t -s|^{2H-2}.
    \label{eq:corr_dBH}
\end{equation}
When $H=1/2$, $\xi_H(t)$ is white uncorrelated noise, and $\left\langle \xi_{1/2}(t) \xi_{1/2}(s)\right\rangle \! = \! \delta(t \! - \! s)$.

Notice that, when the noise strength is constant $a(x)=\theta_H$, Eq.~\eqref{eq:mfBm} reduces to the fBm $B_H(t) \! \equiv \! x|_{a(x) =\theta_H} \! = \! \theta_H \! \int_{0}^t \xi_H(\tau)d\tau$, with correlations
\begin{equation}
    \left\langle B_H(t)B_H(s) \right\rangle = \frac{\theta_H^2}{2} \left(t^{2H} +s^{2H} - \left|t -s\right|^{2H} \right),
    \label{eq:corr_BH}
\end{equation}
such that for $H \!\! > \!\! 1/2$ ($H \!\! < \!\! 1/2$), the increments of $x$ are positively (negatively) correlated with persistent (antipersistent) long-term dependence. 

The cumulant generating functional density of the noise $\xi_H(t)$ (see App.~\ref{app:CGF}) can be written by rescaling its conjugate variable $z(t)\to a(x)z(t)$,
\begin{equation}
\begin{aligned}[b]
    \kappa_H [a(x)z(t)] \!=\!-\frac{1}{2}\Gamma(2H \!\!+ \!\!1)a(x)z(t) \!\prescript{RL}{0}{I}_{t}^{2H-1}\!\left[a(x)z(t)\right] \!,
    \label{eq:kappa_mfBm}
\end{aligned}
\end{equation}
where $\prescript{RL}{0}{I}_{t}^{2H-1}\left[f(x(t), t)\right]$ is the Riemmann-Liouville fractional integral acting over the function $f(x,t) = a(x)z(t)$,
\begin{equation}
	\prescript{RL}{0}{I}_{t}^{2H-1} \!f(t)\! = \! \frac{1}{\Gamma(2H \! - \!1)} \! \int_{0}^{t} (t \! - \! \tau)^{2H-2} f(x(\tau), \tau) d\tau.
	\label{eq:RLFI}
\end{equation}
Integrating in time, we get the cumulant generating functional $\mathcal{K}(z(t)) \! = \! \int_0^t \kappa(z(\tau)) d\tau$, which satisfies the inversion formula,
\begin{equation}
    e^{\mathcal{K}_{H} \left( z \right)} \! = \! e^{\int_0^{t} \kappa_{H} \left( z \right) d\tau} \!\! = \!\! \int \mathcal{D}\xi_H \mathcal{P} \!\left(\xi_H\right) e^{-i \! \int_{0}^t z(\tau)\xi_H(\tau)d\tau}.
    \label{eq:functional_CGF}
\end{equation}

Now, the transition probability of the stochastic process \eqref{eq:mfBm} can be written as a propagator using the Parisi-Sourlas method \cite{parisi1979random},
\begin{equation}
    \mathcal{P}\left( x, t \right.\left| x_{0}, t_{0} \right) \! = \! \left\langle \int\!
    \mathcal{D}x\;\mathcal{J}\!\left(\!\frac{\partial\xi_H}{\partial x}\!\!\right)\delta\!\left[\dot{x} -a(x)\xi_H\right]\!
    \right\rangle_{\!\xi_H},
    \label{eq:P1}
\end{equation}
where $\mathcal{J}\left(\partial\xi_H/\partial x\right)$ is the Jacobian of the transformation between $\xi_H$ and $x$, and $\left\langle \cdot \right\rangle_{\xi_H}$ represents the ensemble average over the noise. The Dirac delta in Eq.~\eqref{eq:P1} filters out the trajectories that do not satisfy the stochastic differential equation \eqref{eq:mfBm}.

Recalling the functional Dirac delta,
\begin{equation}
\begin{aligned}[b]
\delta\!\left[\dot{x} \!- \!a(x)\xi_H \right] = \int \! \frac{\mathcal{D}z}{2\pi} \exp{\!\left(\!i \!\int_0^{t} \!\! z(\tau) \! \left[\dot{x} \! -\! a(x)\xi_H\right] d\tau \!\right)},
\label{eq:functional_diracdelta}
\end{aligned}
\end{equation}
and explicitly writing the ensemble average in terms of the probability measure of the noise $\mathcal{P} \!\left(\xi_H\right) \! \mathcal{D}\xi_H$ and Eq.~\eqref{eq:functional_CGF}, the propagator reads
\begin{equation}
\begin{aligned}[b]
    \mathcal{P}\left(x, t| x_{0},t_{0} \right) \! &= \!\!\! \int \!\! \mathcal{D}x \;\mathcal{J}\!\!\left( \! \frac{\partial\xi_H}{\partial x} \!\!\right)\\
    &\times \!\!\int \! \frac{\mathcal{D}z}{2\pi}e^{\int_0^{t} iz(\tau)\dot{x}(\tau)+\kappa_H[a(x)z(\tau)] d\tau},
    \label{eq:P2}
\end{aligned}  
\end{equation}

The Jacobian of the process is simply given by $\mathcal{J}\!\!\left(\partial {\xi_H}/\partial x \right) \! = \![a(x_0)/a(x)]^{\gamma}$ (see Eq.~\eqref{eq:J3} in App.~\ref{app:Jacobian}). When $\gamma=0$, it reduces to $\mathcal{J}\!\!\left(\partial {\xi_H}/\partial x \right) \! =1$. On the other hand, when $\gamma\neq0$, this term plays an important role for calculating detailed balance, irreversibility and entropy production, since it contributes to the action asymmetry between forward and backward paths \cite{abril2025path}. In particular, for the calculation of the forward propagator, it can be absorbed into the measure $\mathcal{D}x$ \cite{balaji2009universal}, provided that $a(x)$ is a non-vanishing $\mathcal{C}^1$ function--i.e. its first derivative exists and is continuous over its entire domain--. Thus, the path integral can be simplified to
\begin{equation}
    \mathcal{P}\left(x, t| x_{0},t_{0} \right) = \int \mathcal{D}x\int \frac{\mathcal{D}z}{2\pi}e^{\int_0^{t} \mathcal{L}(x,\dot{x},z,\tau) d\tau},
    \label{eq:P3}
\end{equation}
with the Lagrangian,
\begin{equation}
    \mathcal{L}(x,\dot{x},z,\tau) = iz(\tau)\dot{x}(\tau)+\kappa_H[a(x)z(\tau)].
    \label{eq:L}
\end{equation}

The Hamiltonian corresponding to Eq.~\eqref{eq:L} is given by
\begin{equation}
\begin{aligned}[b]
    \mathcal{H}(x, \! p_x, \! \tau) \! &= \!\! \sum_{q\in\{x,z\}} p_q\dot{q} - \mathcal{L}[x,\dot{x},z,\tau] \\
    & = -\frac{1}{2}\Gamma(2H \!\! + \!\! 1)a(x)p_x \! \prescript{RL}{0}{I}_{\tau}^{2H-1} \!\left[a(x) p_x\right],
    \label{eq:H}
\end{aligned}
\end{equation}
with the canonical conjugate momenta $p_x = \partial\mathcal{L}/\partial\dot{x} = i z$ and $p_z \! = \! \partial\mathcal{L}/\partial\dot{z} \! = \! 0$. When $H \! = \! 1/2$ and $a(x) \! = \! \theta_{1/2}$, Eqs.~\eqref{eq:P3} and \eqref{eq:L} lead to the Onsager-Machlup action \cite{onsager1953fluctuations,machlup1953fluctuations}, where the kinetic contribution is local in time and given by $\kappa_{1/2} \! \left( \theta_{1/2} p_x \right) \! = \! \theta_{1/2}^2p_x^{2}/2$.

\section{\label{stationary} Stationary-phase solution of the propagator}

The action
\begin{equation}
    S[\varsigma(t)] \! = \!\!\! \int_0^{t} \!\!\left[-iz\dot{x} \! + \! \frac{1}{2}\Gamma(2H \!\! + \!\! 1)a(x)z \! \prescript{RL}{0}{I}_{\tau}^{2H-1} \!\left[a(x)z\right]\right] \! d\tau,
    \label{eq:action}
\end{equation}
where the phase-space trajectory $\varsigma(t) = [x(t), \! z(t)]$ is quadratic in $z(t)$. Thus, for multiplicative coefficients up to quadratic order, $\mathcal{O}[a(x)]\leq \mathcal{O}(x^2)$, it can be exactly evaluated using a stationary-phase approximation. Expanding around the classical trajectory $\bar{\varsigma}(t)$, such that $\varsigma(t) \! = \! \bar{\varsigma}(t) \! + \! \varrho(t)$, with a small fluctuation $\varrho(t)$ around $\bar{\varsigma}(t)$, the propagator can be rewritten as
\begin{equation}
    \mathcal{P}\left(x, t| x_{0},t_{0} \right) = \mathcal{F}[\varrho(t),\varrho(t_0)]e^{-S[\bar{\varsigma}(t)]},
    \label{eq:P4}
\end{equation}
with $\mathcal{F}[\varrho(t),\varrho(t_0)]$ a normalization constant that does not depend on the classical trajectory.

Extremizing the action at the classical trajectory, we obtain the set of equations
\begin{equation}
    \begin{aligned}
       \frac{\delta S}{\delta z(s)}\bigg|_{\varsigma=\bar{\varsigma}} = 0 
       &= -i\dot{\bar{x}}(s)+a(\bar{x})I_{0^+}(s),
       \\ 
       \frac{\delta S}{\delta x(s)}\bigg|_{\varsigma=\bar{\varsigma}} = 0
       &= i\dot{\bar{z}}(s)+a'(\bar{x})\bar{z}(s)I_{0^+}(s),
    \end{aligned}
    \label{eq:minima}
\end{equation}
with the fractional integral,
\begin{equation}
    I_{0^+}(s) = \int_0^s d\tau a(\bar{x})\bar{z}(\tau)C_H(|\tau-s|).
\label{eq:I0}
\end{equation}
Taking the ratio between Eqs.~\eqref{eq:minima}, we obtain a much simpler linear equation,
\begin{equation}
    \frac{\dot{\bar{z}}(s)}{\bar{z}(s)} = -\frac{a'(\bar{x})}{a(\bar{x})}\dot{\bar{x}}(s),
\end{equation}
or equivalently, $d\ln[\bar{z}(s)] = -d\ln[a(\bar{x})]$. Its solution satisfies
\begin{equation}
    \bar{z}(t)a(\bar{x})=\lambda,
    \label{eq:constant}
\end{equation}
with the constant of motion $\lambda$, which depends on the fixed extremes of the trajectory.

Equation \eqref{eq:constant} enormously simplifies our analysis, since it allows to break the convolutions involving the time correlations $C_H(|t-s|)$. Solving the fractional integral in Eq.~\eqref{eq:I0}, we find
\begin{equation}
    I_{0^+}(s) = \lambda H s^{2H-1}.
\label{eq:I0_sol}
\end{equation}
Now, replacing Eq.~\eqref{eq:I0_sol} into the first equation in \eqref{eq:minima}, we obtain the linear equation,
\begin{equation}
    i\dot{\bar{x}}-\lambda H a(\bar{x})s^{2H-1} = 0.
    \label{eq:minima_local}
\end{equation}
Using the Lamperti transform \cite{lamperti1962semi},
\begin{equation}
	Y[x(t), t] \equiv \int d\tilde{x} \frac{1}{a(\tilde{x})}\bigg|_{\tilde{x}=x(t)},
	\label{eq:lamperti}
\end{equation}
Eq.~\eqref{eq:minima_local} can be easily integrated, yielding a compact expression for the constant of motion
\begin{equation}
    \lambda=2i\frac{Y[\bar{x}(t), t]-Y[\bar{x}_0, t_0]}{t^{2H}}.
    \label{eq:lambda}
\end{equation}
Finally, this last result leads to the stationary action,
\begin{equation}
\begin{aligned}[b]
    S[\bar{\varsigma}(t)] 
    & \! = \! -i\lambda\int_0^{t} \!\! d\tau\frac{\dot{\bar{x}}}{a(\bar{x})} \! + \! \frac{\lambda^2}{2} \!\! \int_0^{t} d\tau \!\! \int_0^{t} \!\!ds \; C_H(|\tau \! - \! s|) \\
    & \! = \! - i\lambda(Y[\bar{x}(t), t]-Y[\bar{x}_0, t_0])+ \frac{\lambda^2t^{2H}}{4}\\
    & \! = \! \frac{(Y[\bar{x}(t), t]-Y[\bar{x}_0, t_0])^2}{t^{2H}}.
\end{aligned}    
    \label{eq:action_}
\end{equation}

Introducing the transform from Eq.~\eqref{eq:lamperti}, allows us to rewrite the transition probability for $x(t)$,
\begin{equation}
	\mathcal{P}\left(x, t| x_{0},t_{0} \right) = \mathcal{J}\!\!\left(\frac{\partial Y}{\partial x}\right) \mathcal{P}\left(Y[x(t),t]|Y[x_{0},t_{0}]\right),
	\label{eq:P6}
\end{equation}
with the Jacobian $\mathcal{J}(\partial Y/\partial x) \! = \! 1/a(x)$, and the Gaussian propagator,
\begin{equation}
	\begin{aligned}[b]
		\mathcal{P}&\left(Y[x(t),t]|Y[x_{0},t_{0}]\right) = \\
		&\frac{1}{\sqrt{\pi t^{2H}}} \exp\left\{-\frac{(Y[\bar{x}(t), t]-Y[\bar{x}_0, t_0])^2}{t^{2H}}\right\}.
	\end{aligned}
	\label{eq:PGauss}
\end{equation}

When $a(x) \! = \! \theta_H$, Eq.~\eqref{eq:P6} retrieves the propagator of the fBm, which was originally derived from the Riemann-Liouville representation \cite{sebastian1995path,calvo2008path}, 
\begin{equation}
	\mathcal{P} \! \left(x, t| x_{0},t_{0} \right) \! = \! \frac{1}{\sqrt{\pi \theta_H^2 t^{2H}}} \exp{\left(-\frac{[x(t)-x_0]^2}{\theta_H^2t^{2H}}\right)}.
	\label{eq:PfBm}
\end{equation}
Furthermore, from Eq.~\eqref{eq:PGauss} we observe that $Y[x(t), \! t] \! \sim \! B_H(t)$, i.e., the transform \eqref{eq:lamperti} yields an fBm with mean $\langle Y[x(t),t]\rangle \! = \! Y[x_0, t_0]$ and variance $\mathrm{Var}(Y[x(t),t]) \! \propto \! t^{2H}$. In particular, when $a(x)=|x|^{\alpha}$, with $\alpha>0$, Eq.~\eqref{eq:PGauss} recovers the result presented in Ref.~\cite{wang2020anomalous}, obtained from a Fokker-Planck perspective.  

\section{\label{confinement} Kinetic equations and confinement}

In this section, we expand our analysis to study the kinetic equation and the confinement dynamics of the process \eqref{eq:mfBm}. For a space-dependent function $\mathcal{V}(x)$, we define the expected value, or Feynman-Kac functional \cite{kac1949distributions},
\begin{align}
    \Psi \left(x, t\right) & \! = \! \langle e^{-\int_0^{t}\mathcal{V}\left[ x(\tau), \tau \right]d\tau} \rangle_x \notag \\
    & \! = \!\! \int_{\mathcal{C}_{f}\left[ t_{0}, t\right]} \!\! \mathcal{D}x \!\! \int \! \frac{\mathcal{D}z}{2\pi} e^{\int_0^{t} \mathcal{L}\left( x,z,\dot{x},\tau \right) d\tau} e^{-\!\int_0^{t}\mathcal{V}(x)d\tau},
    \label{eq:FK}
\end{align}
where $\langle\cdot\rangle_x$ is the average with respect to the stochastic process $x(t)$ and $\mathcal{C}_{f}\left[t_{0}, t \right]$ is the set of all configurations with fixed endpoints $\left( t_{0}, x_{0} \right)$ and $\left( t, x \right)$. The exponential weight penalizes trajectories proportionally to the time spent in regions where $\mathcal{V}(x)$ is large. Thus, for $\mathcal{V}(x) \! \geq \! 0$, it can be interpreted as a space-dependent killing rate. In particular, choosing $\mathcal{V}(x)$ infinite outside a bounded domain enforces absorbing boundary conditions.

A generalized procedure to connect the propagator with a Fokker-Planck-type equation was derived in Ref.~\cite{abril2025path}, by considering a differential increase in the expected value of functional \eqref{eq:FK}. This leads to a partial differential equation (see App.~\ref{app:GFPE} for details)
\begin{equation}
    \left[\frac{\partial}{\partial t} + \mathcal{H}_{eff} \left(x,\partial_x, t \right) + \mathcal{V} \left(x, t \right)\right] \Psi (x, t) = 0,
    \label{eq:kineticFK}
\end{equation}
where $\mathcal{H}_{eff} \equiv \kappa_{H} \left[ \lambda \right]$ is an effective local Hamiltonian obtained with the stationary-phase approximation, and we used $p_x=\partial_x$. Fixing the limits of integration in Eq.~\eqref{eq:FK} collapses all the history of the process into the transition between initial and final points. Notice that $\mathcal{V}(x)$ is not a potential, although Eqs.~\eqref{eq:FK} and \eqref{eq:kineticFK} may give that impression. The inclusion of a potential that correctly reflects the dynamics of a physical system must be implemented directly at the level of the Langevin equation \eqref{eq:mfBm}; consequently, it would yield a different action than the one described by Eq.~\eqref{eq:action}.

\subsection{Unconfined kinetics}

Let us first focus on the case in which the killing rate vanishes, $\mathcal{V}(x) \! = \! 0$. Equation \eqref{eq:kineticFK} describes the probability of finding the process in a particular state $x(t)$ at time $t$, given the initial state $x_{0}$ at time $t_{0}$. Thus, assuming mass-preserving boundary conditions, $\Psi(\infty,t) \! = \! \Psi(-\infty,t)$, the functional obeys the conservation equation,
\begin{equation}
    \frac{d}{dt} \! \int \! \Psi (x, t)\;dx \! = \! - \! \int \mathcal{H}_{eff} \left(x, \partial_{x}\right)\Psi (x, t)\;dx \! = \! 0.
    \label{eq:mass_conservation}
\end{equation}

Using the constraint \eqref{eq:constant}, i.e. $-ip_x a(x) \! = \! \lambda$, we can compute the effective Hamiltonian from \eqref{eq:H},
\begin{equation}
\begin{aligned}[b]
    \mathcal{H}_{eff}(x, \partial_x) 
    & = \lambda^2Ht^{2H-1} = -Ht^{2H-1}[p_xa(x)]^2 \\
    & = -Ht^{2H-1}\partial_x[a(x)\partial_{x}[a(x)(\cdot)]],
\end{aligned}
\label{eq:H_eff}
\end{equation}
where we chose the operation ordering $p_xa(x)$ that implies the symmetric discretization of the Jacobian $\gamma=1/2$. This Hamiltonian is fully local in time, in comparison with Eq.~\eqref{eq:H}. Replacing it into Eq.~\eqref{eq:kineticFK} yields the local kinetic equation,
\begin{equation}
    \frac{\partial}{\partial t}\mathcal{P} -Ht^{2H-1}\partial_x[a(x)\partial_x\left[a(x)\mathcal{P}\right]]= 0,
    \label{eq:kineticP}
\end{equation}
where we used $\mathcal{P} \! \equiv \! \mathcal{P} \! \left( x, t \right.\left| x_{0}, t_{0} \right)$ to shorten the expression. It is easy to check that the propagator \eqref{eq:P6} is a solution of Eq.~\eqref{eq:kineticP}, by plugging it into the partial differential equation. Furthermore, the limiting case of fBm is fully recovered when the noise is additive.

\subsection{Local vs non-local kinetics}

The approximation \eqref{eq:kineticFK} must be applied with caution. One could attempt to construct the kinetic equation using the non-local Hamiltonian \eqref{eq:H}. However, this leads to a non-local kinetic equation,
\begin{equation}
    \frac{\partial \mathcal{P}}{\partial t} = \frac{1}{2}\Gamma(2H+1)\partial_x\left[a(x)\prescript{RL}{0}{I}_{\tau}^{2H-1}\partial_x\left[a(x) \mathcal{P}\right]\right],
    \label{eq:kineticNonlocal}
\end{equation}
Eq.~\eqref{eq:kineticNonlocal} does not represent the stochastic process \eqref{eq:mfBm}, but rather a process of a different nature, governed by a generalized version of the continuous-time random walk (CTRW) \cite{note_CTRW}.

A simple analysis illustrating this situation can be carried out by considering the case of additive noise, $a(x) \! = \! \theta_H$. By definition, this must reduce to an fBm. However, the non-local kinetic equation reads
\begin{equation}
    \frac{\partial \mathcal{P}}{\partial t} = \theta_H^2 \Gamma(2H+1) \prescript{RL}{0}{I}_{\tau}^{2H-1}\left[\partial_x^2\mathcal{P}\right].
    \label{eq:kineticsCTRW}
\end{equation}
The solution, assuming the initial condition $\mathcal{P} \! = \! \delta(x \! - \!x_{0})$, is given by the Fourier transform of a Mittag-Leffler characteristic function,
\begin{equation}
    \mathcal{P} \! = \!\! \int_{-\infty}^{\infty} \!\! e^{-ik(x-x_{0})}E_{2H}\left(-\theta_H^2\Gamma(2H \!\! + \!\!1)k^{2}t^{2H}\right)\frac{dk}{2\pi},
    \label{eq:CTRW}
\end{equation}
with the two-parameter Mittag-Leffler function defined as
\begin{equation}
    E_{a,b}(z)=\sum_{n=0}^{\infty}\frac{z^{n}}{\Gamma(an+b)}.
    \label{eq:MittagLeffler}
\end{equation}
Equations \eqref{eq:kineticsCTRW} and \eqref{eq:CTRW} have been well studied in the literature as models of CTRWs with power-law waiting times \cite{hilfer1995fractional, metzler2000generalized, thiel2014scaled, wei2023time}. In fact, both fBm and CTRW exhibit the same scaling behavior and mean squared displacement, $\langle (x \! - \! x_0)^2 \rangle \!\sim t^{2H}$. However, their higher-order moments (starting from the fourth) differ, since a CTRW with power-law waiting times is not Gaussian. Consequently, identical MSD does not imply identical stochastic dynamics.

\subsection{Absorbing boundary conditions}

Now, we turn on an infinite killing rate to confine the process $x(t)$ similar to an effective infinite potential well, i.e., $\mathcal{V}(x) \! = \! 0$, if $x_{L} \! < \! x \! < \! x_{U}$, and $\mathcal{V}(x) \! \to \! \infty$, otherwise. Note that a constant killing rate $V_{0} \! \geq \! 0$ implies $\Psi(x,t) \!= \! \langle e^{-V_{0}t}\rangle_{x} \! = \! e^{-V_{0} t} \; \mathcal{P}$. Thus, the propagator satisfies Eq.~\eqref{eq:kineticP} when $x_{L} \! < \! x \! < \! x_{U}$, and $\Psi(x,t) \! = \! 0$ outside the interval due to the exponential damping factor.

Using the mass-preserving boundary conditions, the functional is expressed as
\begin{equation}
    \Psi(x,t) \! = \! \frac{e^{-\frac{(Y[\bar{x}(t), t]-Y[\bar{x}_{0}, t_{0}])^2}{t^{2H}}}}{\mathcal{N}_{H}(x_{L}, x_{U}, t)a(x)\sqrt{\pi t^{2H}}},
    \label{eq:kineticPconfinement}
\end{equation}
where the normalization factor is the cumulative distribution function evaluated in the Lamperti transform $Y[x(t),t] \! \equiv \! Y(x)$, that is,
\begin{equation}
    \mathcal{N}_{H}(x_{L}, x_{U}, t) \! = \!\! \int_{(Y(x_L)-Y(x_0))/t^{H}}^{(Y(x_U)-Y(x_0))/t^{H}} \frac{e^{-u^{2}}}{\sqrt{\pi}} du.
    \label{eq:normConfinement}
\end{equation}

The propagator~\eqref{eq:kineticPconfinement} has a different kinetic equation than Eq.~\eqref{eq:kineticP},
\begin{equation}
    \frac{\partial\Psi}{\partial t} -Ht^{2H-1}\partial_x[a(x)\partial_x\left[a(x)\Psi\right]+\mu_{H}(t)\Psi = 0,
    \label{eq:Confinement}
\end{equation}
where $\mu_{H}(t)$ is an induced artificial drift term given by
\begin{equation}
    \mu_{H}(t)=\frac{d}{dt}\ln{\left[\mathcal{N}_{H}(x_{L}, x_{U}, t)\right]}.
    \label{eq:driftConfinement}
\end{equation}

The artificial drift $\mu_{H}(t)$ acts as a compensating term generated by the time dependence of $N_{H}(x_{L}, x_{U},t)$. This induces a redistribution of probability upon the process $x(t)$ colliding with the interval boundaries, rather than a mechanical drift arising from a real potential in the Langevin equation. Thus, confinement does not create a Boltzmann-type equilibrium; it creates a flat quasi-stationary state because the only spatial structure in the propagator is Gaussian broadening, and after normalization, that structure disappears once the width exceeds the size of the box. Furthermore, the probability conservation is not achieved because the original free process conserves mass within the interval, but rather due to the effect of normalization on $N_{H}(x_{L}, x_{U},t)$ at each time.

At last, note that homogeneous diffusion in the Lamperti transform implies uniform asymptotic behavior in the propagator that depends on the Jacobian metric $a^{-1}(x)$. Thus, the confined multiplicative process naturally accumulates in regions of lower noise amplitude.

\section{\label{conclusions} Conclusions}

Our results provide insights into the construction of a general path-integral representation for the transition probability of heterogeneous diffusion processes driven by fractional Gaussian noise. Using a stationary-phase approximation, we simplify the time-correlated action into a quadratic function of the Lamperti transform of the process. This leads to a Gaussian propagator, which recovers previous findings in the literature \cite{wang2020anomalous}, and in the additive limiting case, retrieves the results based on the Riemann-Liouville formulation of fBm~\cite{calvo2008path,sebastian1995path}, building a bridge with the Langevin representation.

Based on the Feynman-Kac functional~\cite{kac1949distributions,abril2025path}, we study the effects of subordinating the process to a killing rate, and derive a general procedure to construct kinetic equations in terms of the effective local Hamiltonians obtained after the stationary-phase approximation. In particular, when the killing rate is infinite, the process describes absorbing boundary conditions. We show that the interplay between multiplicative diffusion and confinement induces an effective drift term in the kinetic equation, leading to an accumulation of probability in regions of lower noise amplitude.

We envision that the framework presented in this paper can be extended to more general non-Markovian dynamics and noise structures, since the use of cumulant generating functionals allows for systematic simplification of the actions under stationary-phase approximations, suggesting a broader applicability of the method. In this work, we developed on the use of the Feynman-Kac functional for local Hamiltonians. However, it should also be possible to take an alternative route by generalizing Feynman-Kac, in a similar manner to Refs. \cite{turgeman2009fractional,carmi2010distributions,carmi2011fractional}, to work with non-local Hamiltonians.

\section{Acknowledgements}
We are grateful to R. Metzler and E. Barkai for fruitful discussions and valuable comments that helped clarify several aspects of this work.

\bibliography{apssamp}

\appendix
\section{\label{app:CGF} Cumulant generating function of the fractional Gaussian noise}

Recalling that the differential noise increments of the fBm are given by $\xi_ {H_{\mu +1}}\Delta t \equiv B_{H_{\mu +1}}-B_{H_{\mu}}$, the covariance between two increments can be obtained from Eq.~\eqref{eq:corr_BH},
\begin{equation}
\begin{aligned}[b]
    C_{H_{\mu\nu}} 
    & \!\! = \! \langle \xi_{H_\mu}\Delta t \; \xi_{H_\nu}\Delta t\rangle \\
    & \!\! = \! \frac{1}{2} \!\left[|\mu \! - \! \nu \! + \! 1|^{2H} \! + \! |\mu \! - \! \nu \! - \! 1|^{2H} \! - \! 2|\mu \! -\! \nu|^{2H}\right].
    \label{eq:cov_increments}
\end{aligned}    
\end{equation}
Thus, the covariance matrix $\Sigma_{H}$ of $M$ time-correlated increments has elements defined by Eq.~\eqref{eq:cov_increments} and diagonal equal to 1,
\begin{equation}
\left[ \Sigma_{H}\right]_{\mu, \nu} = 
\begin{cases}
      1              & \text{, if } \mu=\nu \\
      C_{H_{\mu\nu}} & \text{, if } \mu\neq\nu \\
\end{cases}
\end{equation}
In particular, for $H \! = \! 1/2$, the matrix reduces to the identity $\Sigma_{1/2} \! = \! \mathbb{I}$, since $C_{1/2_{\mu\nu}}=0$ for $\mu\neq\nu$, which describes $M$ independent Gaussian variables.

Since the increments of the fBm are a stationary Gaussian sequence, centered around zero, the cumulant generating function of the process can be expressed as
\begin{equation}
    \mathcal{K}_{H}(z) = -\frac{1}{2}z^{\text{T}}\Sigma_{H}z = -\frac{1}{2}\sum_{\mu,\nu=0}^{M} z_{\mu}C_{H_{\mu\nu}} z_{\nu},
    \label{eq:CGF_discrete}
\end{equation}
where $z$ is the canonical conjugate variable of the noise $\xi_{H}$. 
In the continuous-time limit, Eq.~\eqref{eq:CGF_discrete} is extended to the cumulant generating functional
\begin{align}
    \mathcal{K}_H\left( z \right) & = -\frac{1}{2} \int_{0}^{t} \int_{0}^{t} z(\tau) \; C_{H}\left( |\tau-s| \right) \; z(s) \; d\tau ds \notag \\
    & = -\int_{0}^t\;z(\tau)\left(C_{H} \ast z\right)(\tau)d\tau,
    \label{eq:CGF_continuous}
\end{align}
which defines a density functional $\kappa_H \! \left(z \right)$, such that $\mathcal{K}_H(z) \! = \! \int_{0}^{t} \kappa_H[z(\tau)]d\tau$. Using the Riemann-Liouville fractional integral \eqref{eq:RLFI}, 
the density functional reads
\begin{equation}
    \kappa_H \left( z \right) = -\frac{1}{2}\Gamma(2H+1)z(t)  \prescript{RL}{0}{I}_{t}^{2H-1}[z(t)].
    \label{eq:CGF_density}
\end{equation}

\section{\label{app:Jacobian}Construction of the Jacobian}
The Jacobian of the stochastic process \eqref{eq:mfBm} can be constructed by following the standard procedure presented in Ref~\cite{hanggi1989path}. Here, we summarize the most relevant steps for this construction using a general prescription for the evaluation of the multiplicative term. Assuming the infinitesimal time $\varepsilon = (t-t_0)/N$, with $N\to\infty$, Eq.~\eqref{eq:mfBm} can be discretized as
\begin{equation}
    \frac{x_n - x_{n-1}}{\varepsilon} = a(\tilde{x}_n)\xi_{H_n},
    \label{eq:x_n'}
\end{equation}
with $x_n\equiv x(t_n)$, $\xi_{H_n}\equiv \xi_H(t_n)$, $t_n\equiv t_0+n\varepsilon$ and $\tilde{x}_n \equiv \gamma x_n+(1-\gamma)x_{n-1}$, with $0\leq\gamma\leq1$ a parameter associated with the prescription (0: Itô \cite{ito1950stochastic}, 1/2: Stratonovich \cite{stratonovich1966new}, 1: Hänggi \cite{hanggi1982stochastic}). 

Summing Eq.~\eqref{eq:x_n'} from $i=0$ to $i=n$, we obtain
\begin{equation}
    x_n = x_0 + \varepsilon \sum_{i=0}^n a(\tilde{x}_i)\xi_{H_i},
    \label{eq:x_n}
\end{equation}
which allows us to calculate the Jacobian,
\begin{equation}
    \mathcal{J}_n\!\left(\!\frac{\partial x_n}{\partial \xi_{H_n}}\!\right) = \left|\frac{\partial x_n}{\partial \xi_{H_n}}\right| = \frac{\varepsilon|a(\tilde{x}_n)|}{1-\varepsilon \gamma a'(\tilde{x}_n)\xi_{H_n}},
    \label{eq:J_n}
\end{equation}
with $a'(\tilde{x}_n) \equiv da(\tilde{x}_n)/dx_n$.

Now, from Eq.~\eqref{eq:J_n} we can define,
\begin{equation}
\begin{aligned}[b]
    \mathcal{J}&\!\left(\!\frac{\partial\xi_H}{\partial x}\!\!\right)=\prod_{i=1}^N \mathcal{J}_i^{-1}\!\left(\!\frac{\partial x_i}{\partial \xi_{H_i}}\!\right) \\
    & = \varepsilon^{-N}\prod_{i=1}^N|a(\tilde{x}_i)|^{-1} \left(1-\varepsilon\gamma a'(\tilde{x}_i)\xi_{H_i}\right)\\
    & = \varepsilon^{-N}\prod_{i=1}^N|a(\tilde{x}_i)|^{-1} \exp\left(\sum_{i=1}^N-\varepsilon\gamma a'(\tilde{x}_i)\xi_{H_i}\right) \\
    & = \varepsilon^{-N}\prod_{i=1}^N|a(\tilde{x}_i)|^{-1}\exp\left(\sum_{i=1}^N-\varepsilon\gamma\frac{a'(\tilde{x}_i)}{a(\tilde{x}_i)} \frac{x_i - x_{i-1}}{\varepsilon}\right),
    \label{eq:J1}
\end{aligned}   
\end{equation}
where we used the Matrix identity $\text{Det}[\mathbb{I}-\mathbb{M}]=\exp\left\{\text{Tr}[\ln(\mathbb{I}-\mathbb{M})]\right\} = \exp\left\{\text{Tr}[-\mathbb{M} - \mathbb{M}^2/2-...]\right\}$, and truncated up to first order.

The term $\varepsilon^{-N}$ in Eq.~\eqref{eq:J1} can be directly absorbed by the integration measure $\mathcal{D}z/2\pi$. On the other hand, the measure $\mathcal{D}x(t)$ can be redefined as
\begin{equation}
\begin{aligned}[b]
    \mathcal{D}x(t)& =\lim_{N\to\infty}\prod_{i=1}^Ndx_i|a(\tilde{x}_i)|^{-1} \\
    &=\lim_{N\to\infty}\left(\prod_{i=1}^Ndx_i\right)\exp\left(-\varepsilon^{-1}\int_0^tds\ln|a(x(s))|\right).
\end{aligned}
\end{equation}

Thus, the only term that is left out in the Jacobian is
\begin{equation}
\begin{aligned}[b]
    \mathcal{J}&\!\left(\!\frac{\partial\xi_H}{\partial x}\!\!\right)= \prod_{i=1}^N\exp\left(\sum_{i=1}^N-\varepsilon\gamma \frac{a'(\tilde{x}_i)}{a(\tilde{x}_i)} \frac{x_i - x_{i-1}}{\varepsilon}\right),
    \label{eq:J2}
\end{aligned}
\end{equation}
which in the limit $N\to\infty$ reads
\begin{align}
	\mathcal{J}\!\left(\!\frac{\partial\xi_H}{\partial x}\!\!\right)
	& = \exp\left[-\gamma\int_0^t\frac{a'(x)}{a(x)}\dot{x}(\tau)d\tau\right] \notag \\
	& = \exp\left[-\gamma\ln\left(\frac{a(x)}{a(x_0)}\right) \right] = \left[\frac{a(x_0)}{a(x)}\right]^{\gamma}.
    \label{eq:J3}
\end{align}

\begin{widetext}
\section{\label{app:GFPE}Generalized Kinetic equation}

The Feynman-Kac functional~\eqref{eq:FK} is defined as
\begin{equation}
    \Psi \left(x, t\right)
     = \int_{\mathcal{C}_{f}\left[ t_{0}, t\right]}\mathcal{D}x \; \psi\!\left(x, t \right) \int \frac{\mathcal{D}z}{2\pi} e^{-\mathcal{S}}.
    \label{eq:feynman_kac_1}
\end{equation}
The weight
\begin{equation}
\psi \left(x, t\right)
     = e^{-\int_0^{t}\mathcal{V}\left[ x(\tau), \tau \right]d\tau},
    \label{eq:feynman_kac_3}
\end{equation}
satisfies the differential equation,
\begin{equation}
    \frac{d}{dt}\psi \! \left(x,t \right) = -\mathcal{V}\left[x(t), t \right] \psi \! \left(x, t \right).
    \label{eq:feynman_kac_4}
\end{equation}

Considering an infinitesimal time increment $\varepsilon>0$ in the Feynman-Kac functional, Eq.~\eqref{eq:feynman_kac_3} is divided into $M+1$ integrals, with $t_{M+1}=t+\varepsilon$,

\begin{equation}
\begin{aligned}[b]
    \Psi\left(x_{M+1}, t_{M+1}\right) &= \int \left(\prod_{\mu=1}^{M}\int\int \psi\left(x_{\mu},t_{\mu}\right) dx_{\mu+1} d\left(\frac{z_{\mu}}{2\pi}\right)\right) \psi\left( x_{M+1}, t_{M+1}\right) e^{\left(-\sum_{\mu=1}^{M} \mathcal{S}_{\mu}\right)} \; d\left(\frac{z_{M+1}}{2\pi}\right) \\
    & = \int \left(\prod_{\mu=1}^{M}\int\int \psi\left(x_{\mu},t_{\mu}\right) e^{-\mathcal{S}_{\mu}} dx_{\mu +1} d\left(\frac{z_{\mu}}{2\pi}\right)\right) \psi\left( x_{M+1}, t_{M+1}\right) \; d\left(\frac{z_{M+1}}{2\pi}\right) \\
    & = \int \left(\int \Psi\left( x_{M}, t_{M}\right) e^{-\mathcal{S}_{M}} \; dx_{M+1}\right) \psi\left( x_{M+1}, t_{M+1}\right) \; d\left(\frac{z_{M+1}}{2\pi}\right) \\
    & =  \int \Psi\left( x_{M}, t_{M}\right) \; \psi\left( x_{M+1}, t_{M+1}\right) \int e^{-\mathcal{S}_{M}} d\left(\frac{z_{M+1}}{2\pi}\right)\;dx_{M+1},
    \label{eq:feynman_kac_6}
\end{aligned}
\end{equation}
where $\mathcal{S}_{\mu}$ is the action between times $t_{\mu -1}$ and $t_{\mu}$, $\psi\left( x_{\mu}, t_{\mu}\right) = \psi\left( x_{\mu}, t_{\mu} \right.\left| x_{\mu-1}, t_{\mu-1}\right)$, for all $\mu \in \{ 1, 2, \dots, M+1\}$, and the additional integrals over the conjugate variable of the noise $z(t)$ are due to the fixed endpoints, which means there is always an additional integral for each degree of freedom in the system.

At first order in $\varepsilon$, we have that
\begin{equation}
\begin{aligned}[b]
    \mathcal{I}_{M} & = \int e^{-\mathcal{S}_{M}} \frac{dz}{2\pi} = \int e^{-i z \left( x_{M+1} - x_{M}\right) -\varepsilon \kappa_{H} \left[a(x_{M+1})z \right]} \frac{dz}{2\pi} \\
    & = \int e^{-i z\left(x_{M+1} - x_{M}\right)}\left[ 1 - \varepsilon \kappa_{H} \left[a(x_{M+1})z \right] \right] \frac{dz}{2\pi} \\
    & = \left( 1 - \varepsilon \kappa_{H} \left[ \lambda \right] \right)\int e^{-i z\left(x_{M+1} - x_{M}\right)} \frac{dz}{2\pi} \\
    & = \delta\left(x_{M+1} - x_{M}\right) \left( 1 - \varepsilon \kappa_{H} \left[ \lambda \right] \right),
    \label{eq:feynman_kac_7}
\end{aligned}
\end{equation}
since the integral of $i z \dot{x}$ is discretized as $i z_{\mu}\left( x_{\mu} - x_{\mu -1}\right)$, and using the constraint \eqref{eq:constant}, i.e., $-ip_xa(x)=\lambda$. Thus, from Eqs.~\eqref{eq:feynman_kac_4} and \eqref{eq:feynman_kac_7}, we have
\begin{equation}
\begin{aligned}
    \psi\left(x_{M+1}, t_{M+1}\right)\mathcal{I}_{M} & = \psi\left(x_{M+1}, t_{M+1} \right.\left| x_{M}, t_{M}\right)\mathcal{I}_{M} \\
    & = \left(\psi\left(x_{M}, t_{M} \right.\left| x_{M}, t_{M} \right) + \varepsilon \left.\frac{d}{dt_{M+1}}\psi\left(x_{M+1}, t_{M+1} \right.\left| x_{M}, t_{M}\right) \right|_{\varepsilon=0} \right) \mathcal{I}_{M}\\
    & = \left[1 - \varepsilon \mathcal{V}\left(x_{M}, t_{M}\right) \right] \; \delta\left(x_{M+1} - x_{M}\right) \left( 1 - \varepsilon \kappa_{H} \left[ \lambda \right] \right) \\
    & = \left[ 1 - \varepsilon \kappa_{H} \left[ \lambda \right] - \varepsilon \mathcal{V}\left(x_{M}, t_{M}\right)\right]\delta\left(x_{M+1} - x_{M}\right) + \mathcal{O}\left( \varepsilon^{2}\right).
    \label{eq:feynman_kac_8}
\end{aligned}
\end{equation}

Replacing Eq.~\eqref{eq:feynman_kac_8} in Eq.~\eqref{eq:feynman_kac_6}, we conclude
\begin{equation}
    \Psi\left(x_{M+1}, t_{M+1}\right) = \left[ 1 - \varepsilon \kappa_{H} \left[ \lambda \right] - \varepsilon \mathcal{V}\left(x_{M}, t_{M}\right)\right] \Psi\left( x_{M}, t_{M}\right).
    \label{eq:feynman_kac_9}
\end{equation}
Finally, taking the limit $\varepsilon\to0^{+}$, Eq.~\eqref{eq:feynman_kac_9} reduces to the Eq. \eqref{eq:FK}.

\end{widetext}

\end{document}